\documentclass[12pt]{article}
\usepackage{amssymb}
\usepackage{graphicx}

\def\be{\begin{equation}}
\def\ee{\end{equation}}

\begin{document}
\begin{center}
{\Large\bf Critical behavior of the compact $3d$ $U(1)$ gauge theory
on isotropic lattices}\\ 
\end{center}
\vskip 0.3cm
\centerline{O. Borisenko$^*$}
\vskip 0.3cm
\centerline{\sl Bogolyubov Institute for Theoretical Physics,}
\centerline{\sl National Academy of Sciences of Ukraine,}
\centerline{\sl 03680 Kiev, Ukraine}
\vskip 0.3cm
\centerline{R. Fiore$^{**}$}
\vskip 0.3cm
\centerline{\sl Dipartimento di Fisica, Universit\`a della Calabria,}
\centerline{\sl and Istituto Nazionale di Fisica Nucleare, Gruppo collegato
di Cosenza}
\centerline{\sl I-87036 Arcavacata di Rende, Cosenza, Italy}
\vskip 0.3cm
\centerline{M. Gravina$^{***}$}
\vskip 0.3cm
\centerline{\sl Laboratoire de Physique Th\'eorique, Universit\'e de Paris-Sud 11,
B\^atiment 210}
\centerline{\sl 91405 Orsay Cedex France}
\vskip 0.3cm
\centerline{A. Papa$^{**}$}
\vskip 0.3cm
\centerline{\sl Dipartimento di Fisica, Universit\`a della Calabria,}
\centerline{\sl and Istituto Nazionale di Fisica Nucleare, Gruppo collegato
di Cosenza}
\centerline{\sl I-87036 Arcavacata di Rende, Cosenza, Italy}
\vskip 0.6cm

\begin{abstract}
We report on the computation of the critical point of the deconfinement phase 
transition, critical indices and the string tension in the compact three dimensional 
$U(1)$ lattice gauge theory at finite temperatures. The critical indices govern 
the behavior across the deconfinement phase transition in the pure gauge $U(1)$ 
model and are generally expected to coincide with the critical indices of the 
two-dimensional $XY$ model. We studied numerically the $U(1)$ model for $N_t=8$ on 
lattices with spatial extension ranging from $L=32$ to $L=256$. Our determination
of the infinite volume critical point on the lattice with $N_t=8$ differs substantially 
from the pseudo-critical coupling at $L=32$, found earlier in the literature and
implicitly assumed as the onset value of the deconfined phase.
The critical index $\nu$ computed from the scaling of the pseudo-critical couplings 
with the extension of the spatial lattice agrees well with the $XY$ value $\nu =1/2$. 
On the other hand, the index $\eta$ shows large deviation from the expected universal 
value. The possible reasons of such behavior are discussed in details. 
\end{abstract}

\vfill
\hrule
\vspace{0.3cm}
{\it e-mail addresses}: 

$^*$oleg@bitp.kiev.ua, $^{**}$fiore,papa@cs.infn.it,
$^{***}$Mario.Gravina@th.u-psud.fr

\newpage

\section{Introduction}

In this article we continue our exploration of the deconfinement phase 
transition in the three-dimensional ($3d$) $U(1)$ lattice gauge theory (LGT)
started in Ref.~\cite{beta_szero}. The partition function of the compact version 
of this model can be written as
\begin{equation}
Z(\beta_t,\beta_s) = \int_0^{2\pi}\prod_{x\in\Lambda}\: \prod_{n=0}^2
\frac{d\omega_n (x)}{2\pi} \ \exp{S[\omega]} \ ,
\label{PTdef}
\end{equation}
where $\Lambda$ is an $L^2\times N_t$ lattice, $S$ is the Wilson action 
\begin{equation}
 S[\omega] = \beta_s\sum_{p_s} \cos\omega (p_s) + 
\beta_t\sum_{p_t} \cos\omega (p_t) 
\label{wilsonaction} 
\end{equation}
and sums run over all space-like ($p_s$) and time-like ($p_t$) plaquettes.
The plaquette angles $\omega(p)$ are defined in the standard way. The anisotropic 
couplings $\beta_t$ and $\beta_s$ are defined in Ref.~\cite{beta_szero}. 
Since we study the theory at finite temperature, periodic boundary conditions 
in the temporal direction are imposed on the gauge fields.

Let us recapitulate briefly what is known and/or expected about the critical 
behavior of the $3d$ $U(1)$ LGT at finite temperature. At zero temperature 
the theory is confining at all values of the bare coupling constant~\cite{polyakov},
while at finite temperature the theory undergoes a deconfinement phase transition. 
It is well known that the partition function of the $3d$ $U(1)$ LGT in the Villain 
formulation coincides with that of the $2d$ $XY$ model in the leading order of 
the high-temperature expansion~\cite{parga}. When combined with the universality 
conjecture by Svetitsky-Yaffe~\cite{svetitsky}, this result makes one to 
conclude that the deconfinement phase transition belongs to the universality class 
of the $2d$ $XY$ model, which is known to have Berezinskii-Kosterlitz-Thouless (BKT) 
phase transition of infinite order~\cite{Berezinsky:1970fr,Kosterlitz:1973xp}. 
It is therefore generally expected the critical behavior of the $3d$ $U(1)$ LGT 
to coincide with that of the $XY$ model. In particular, one might expect the critical 
behavior of the Polyakov loop correlation function $\Gamma (R)$ to be governed 
by the following expressions
\begin{equation}
\Gamma (R) \ \asymp \ \frac{1}{R^{\eta (T)}} \ ,
\label{PLhight}
\end{equation}
for $\beta \geq \beta_c$ and 
\begin{equation}
\Gamma (R) \ \asymp \ \exp \left [ -R/\xi (t)  \right ] \ ,
\label{PLlowt}
\end{equation}
for $\beta < \beta_c$, $t=\beta_c/\beta -1$.
Here, $R\gg 1$ is the distance between test charges, $T$ is the temperature
and $\xi \sim 
e^{bt^{-\nu}}$ is the correlation length. Such behavior of $\xi$ defines the 
so-called {\it essential scaling}. The critical indices $\eta (T)$ and $\nu$ 
are known from the renormalization-group (RG) analysis of the $XY$ 
model: 
$\eta (T_c) =1/4$ and $\nu=1/2$, where $T_c$ is the BKT critical point. 
Therefore, the critical indices $\eta$ and $\nu$ should be the same in the 
finite-temperature $U(1)$ model if the Svetitsky-Yaffe conjecture holds in 
this case.

The renormalization-group calculations of the RG flow, presented in Ref.~\cite{svetitsky}, 
gave support to the BKT scaling scenario. However, the critical indices have not been 
computed. The direct numerical check of 
these predictions was performed on lattices $L^2\times N_t$ with $L=16, 
32$ and $N_t=4,6,8$ in Ref.~\cite{mcfinitet}. Though the authors of 
Ref.~\cite{mcfinitet} confirm the expected BKT nature of the phase transition, 
the reported critical index is almost three times that predicted for the $XY$ 
model, $\eta (T_c) \approx 0.78$. 
More recent numerical simulations of Ref.~\cite{chernodub} have been mostly 
concentrated on the study of the properties of the high-temperature phase. 
What is important for us here is the derivation of the critical point in 
Refs.~\cite{mcfinitet,chernodub}. In these papers it was found that, 
for the isotropic lattice $\beta_s=\beta_t=\beta$ with $L=32$ and $N_t=8$, 
the pseudo-critical point is $\beta_{pc}=2.30(2)$ for Ref.~\cite{mcfinitet} and 
$\beta_c\approx 2.346(2)$ for Ref.~\cite{chernodub}. Values of $\beta$ above these 
values were taken implicitly as belonging to the deconfined phase.
We shall comment on this derivation later since our result for the infinite volume 
critical coupling differs essentially for this choice of $N_t$.

In our previous paper~\cite{beta_szero} we have studied the model on 
extremely anisotropic lattice with $\beta_s=0$. In this limit the model exhibits 
the deconfinement phase transition which gives the possibility to study the critical 
behavior. We presented simple analytical consideration which showed 
that in the limits of both small and large $\beta_t$ such anisotropic model 
reduces to the $2d$ $XY$ model with some effective couplings. Then we performed 
numerical simulations of the effective spin model for the Polyakov loop which can be 
exactly computed in the limit $\beta_s=0$. We used lattices with $N_t=1,4,8$ and 
with the spatial extent $L\in [64,256]$. Our main goal was to determine the critical 
index $\eta$ supposing that the scaling known from the study of the $XY$ model holds 
also in our case. The main conclusion of our investigation was that the value of 
the index $\eta$ is well compatible with the $XY$ value. We may thus assume that 
at least in the limit $\beta_s=0$ the $3d$ $U(1)$ LGT does belong to the universality 
class of the $XY$ model. 

Encouraged by these findings we have decided to simulate directly the isotropic model 
on the lattice with $N_t=8$. In this paper we present the results of these simulations 
for a number of different quantities. Our general strategy is essentially the same 
as in the previous paper. Namely, we postulate that the scaling laws of the $XY$ model 
can be used to study the critical behavior of the gauge model. We believe that 
the information gathered so far allows for such an approach to be trustworthy. 
Nevertheless, in doing so we have encountered certain surprises. First of all, 
the infinite volume critical coupling turned out to be essentially higher than 
the values for the pseudo-critical couplings reported in 
Refs.~\cite{mcfinitet,chernodub}. As a consequence, the values of $\beta$ used 
in Ref.~\cite{chernodub} to study the deconfinement phase lie well inside the 
confinement phase when the thermodynamic limit is considered.
Secondly, the index $\nu$ extracted from the scaling of the 
pseudo-critical couplings with $L$ does agree well with the expected $XY$ value 
$\nu=1/2$. 
However, the index $\eta$ was found to be strikingly different from 
the $XY$ value, namely $\eta \approx 0.50$. While the value $\eta \approx 0.78$ 
obtained in Ref.~\cite{mcfinitet} could, in principle be attributed to rather small 
lattices used, $L=32$, and to an incorrect location of the critical point, 
our result is almost insensitive to varying the spatial extent if $L$ is large 
enough.  

This paper is organized as follows. In the next section we describe briefly our 
numerical procedure. The result of simulations are presented in the Section~3. 
Conclusions and discussion are given in the Section~4.

\section{Numerical set-up}
\label{setup}

With the aim of calculating the critical indices and then identifying the 
universality class of the $3d$ $U(1)$ LGT, we simulated the system on lattices of the type 
$L^2 \times N_t$, with $N_t=8$ fixed and $L$ increasing towards the thermodynamic
limit. In the adopted Monte Carlo algorithm a sweep consisted in a mixture of one 
Metropolis update and five microcanonical steps. Measurements were taken every 10 
sweeps in order to reduce the autocorrelation and the typical statistics per run was 
about 100k. The error analysis was performed by the jackknife method over bins at
different blocking levels.

\begin{figure}[tb]
\centering
\includegraphics[width=5.1cm]{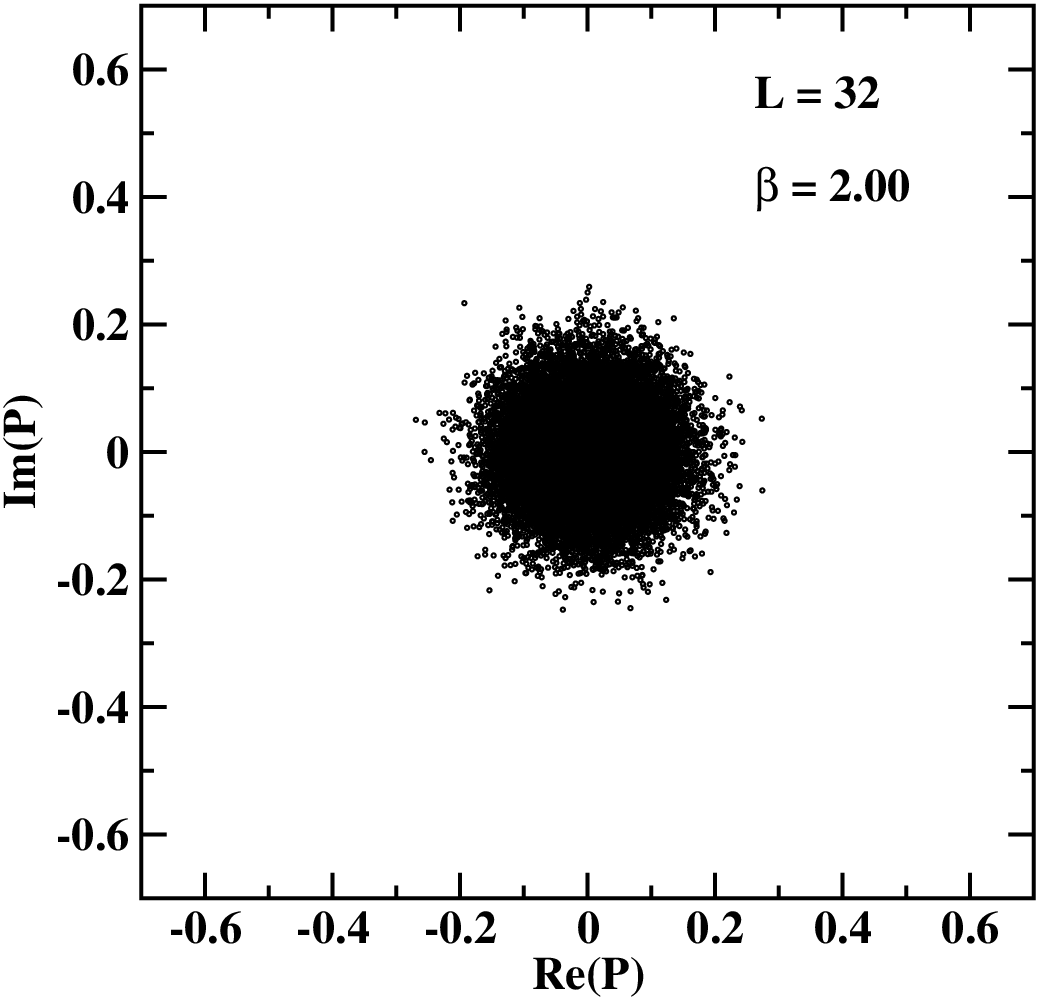}
\hspace{0.1cm}
\includegraphics[width=5.1cm]{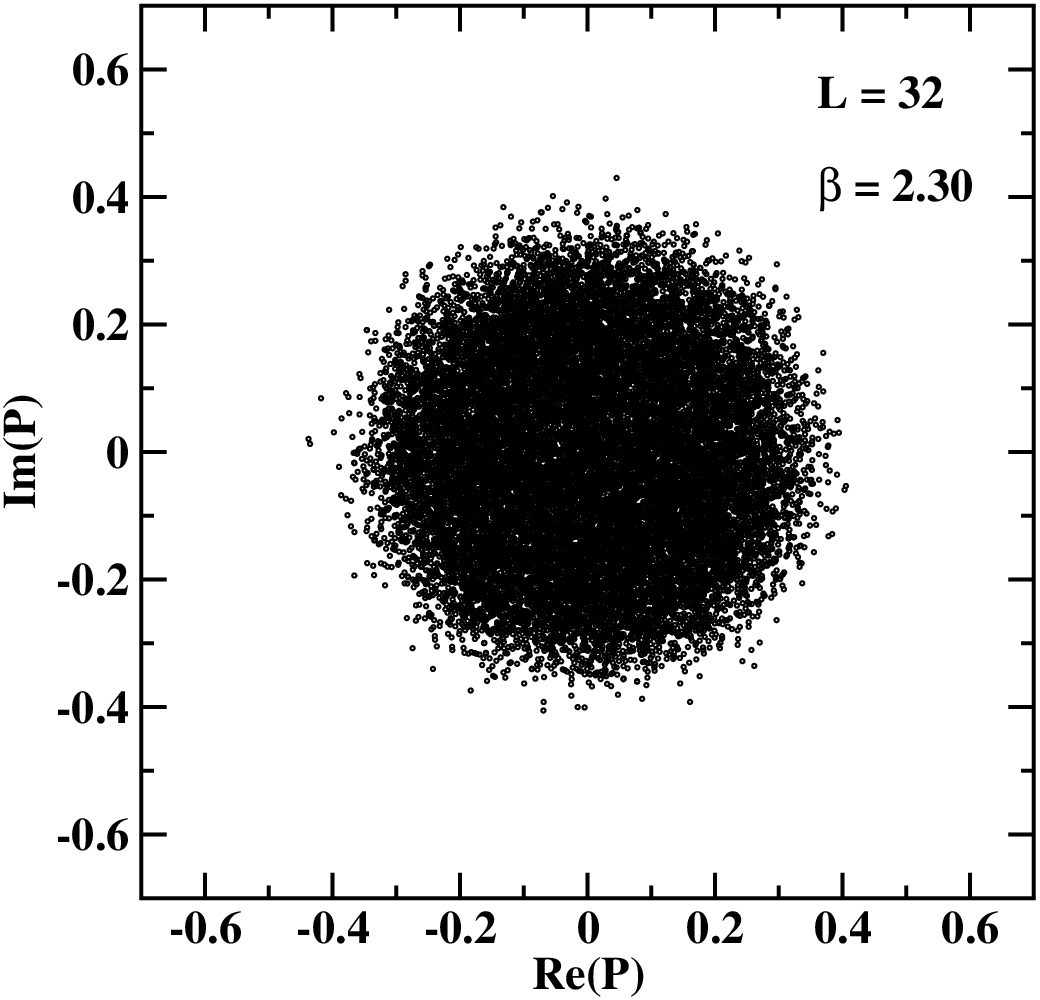}
\hspace{0.1cm}
\includegraphics[width=5.1cm]{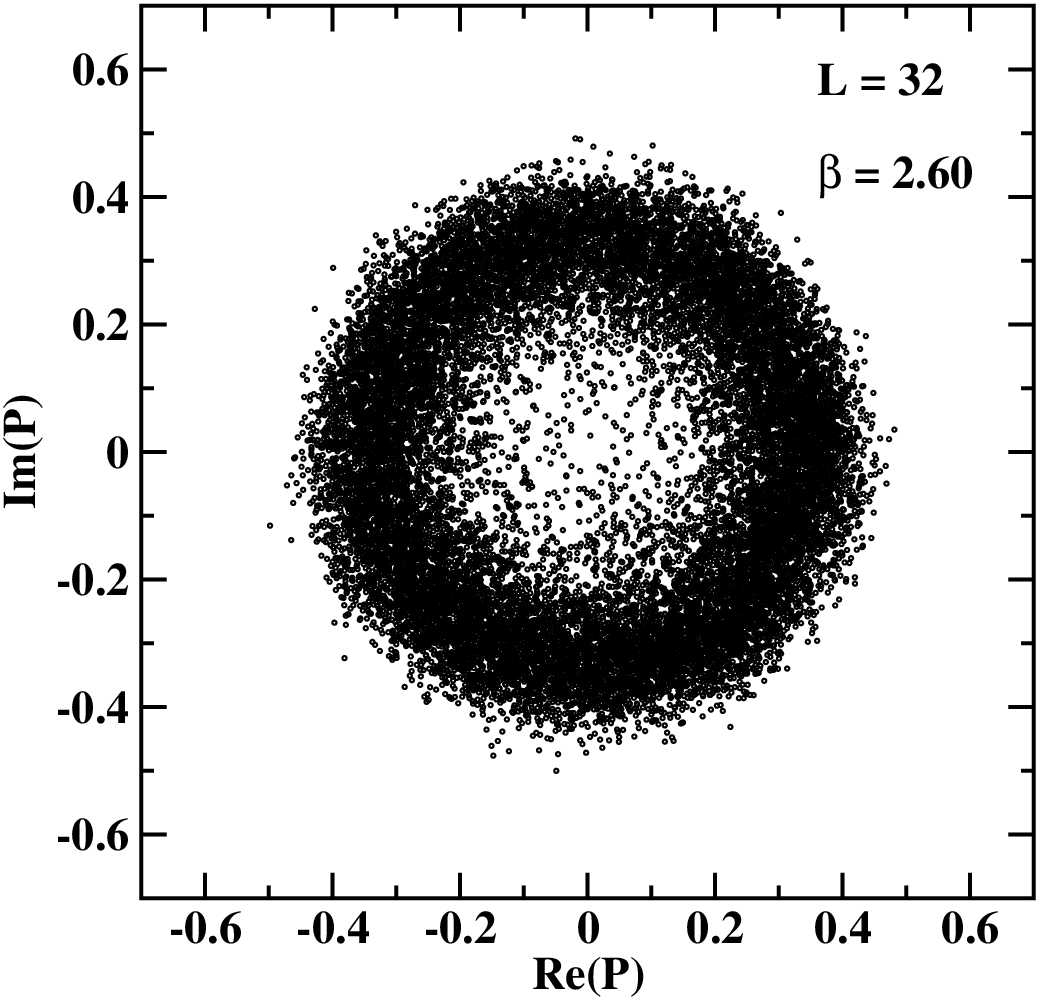}
\caption[]{Scatter plot of the Polyakov loop for $\beta=2.00, 2.30, 2.60$ on the 
$32^2\times 8$ lattice.}
\label{scatter}
\end{figure}

The observable used as a probe of the two phases of the finite temperature $3d$ $U(1)$ LGT
is the Polyakov loop, defined as
\begin{equation}
P(\vec{x})=\prod_t U_0(\vec{x},t) \quad ,
\end{equation}
where $U_0(\vec{x},t)$ is the temporal link attached at the spatial point $\vec{x}$.
The effective theory for the Polyakov loop is two-dimensional and possesses global
$U(1)$ symmetry. Since the global symmetry cannot be broken spontaneously in two dimensions 
owing to the Mermin-Wagner-Coleman theorem the expectation value of the Polyakov loop 
vanishes in the thermodynamic limit. On a finite lattice 
$\langle \sum_{\vec{x}} P(\vec{x}) \rangle =0$ due to $U(1)$ symmetry (if the boundary conditions 
used preserve the symmetry). This is confirmed by the numerical analysis on the periodic lattice: 
in the confined (small $\beta$) phase 
the values taken by the Polyakov loop in a typical Monte Carlo ensemble scatter around 
the origin of the complex plane forming a uniform cloud, whereas in the deconfined
(high $\beta$) phase they distribute on a ring, the thermal average being 
equal to zero in both cases (see Fig.~\ref{scatter} for an example of this
behavior in the case $L=32$, where the transition occurs at $\beta$=2.346(2),
according to Ref.~\cite{chernodub}).
What really feels the transition is then the absolute value of $P$, which has been 
chosen to be the order parameter in this work. It is worth to stress that this kind 
of dynamics is the same presented by the spin magnetization in the $2d$ $XY$ model.

\section{Results at $\beta_s=\beta_t$}
\label{results}

\begin{figure}[tb]
\centering
\includegraphics[width=15cm]{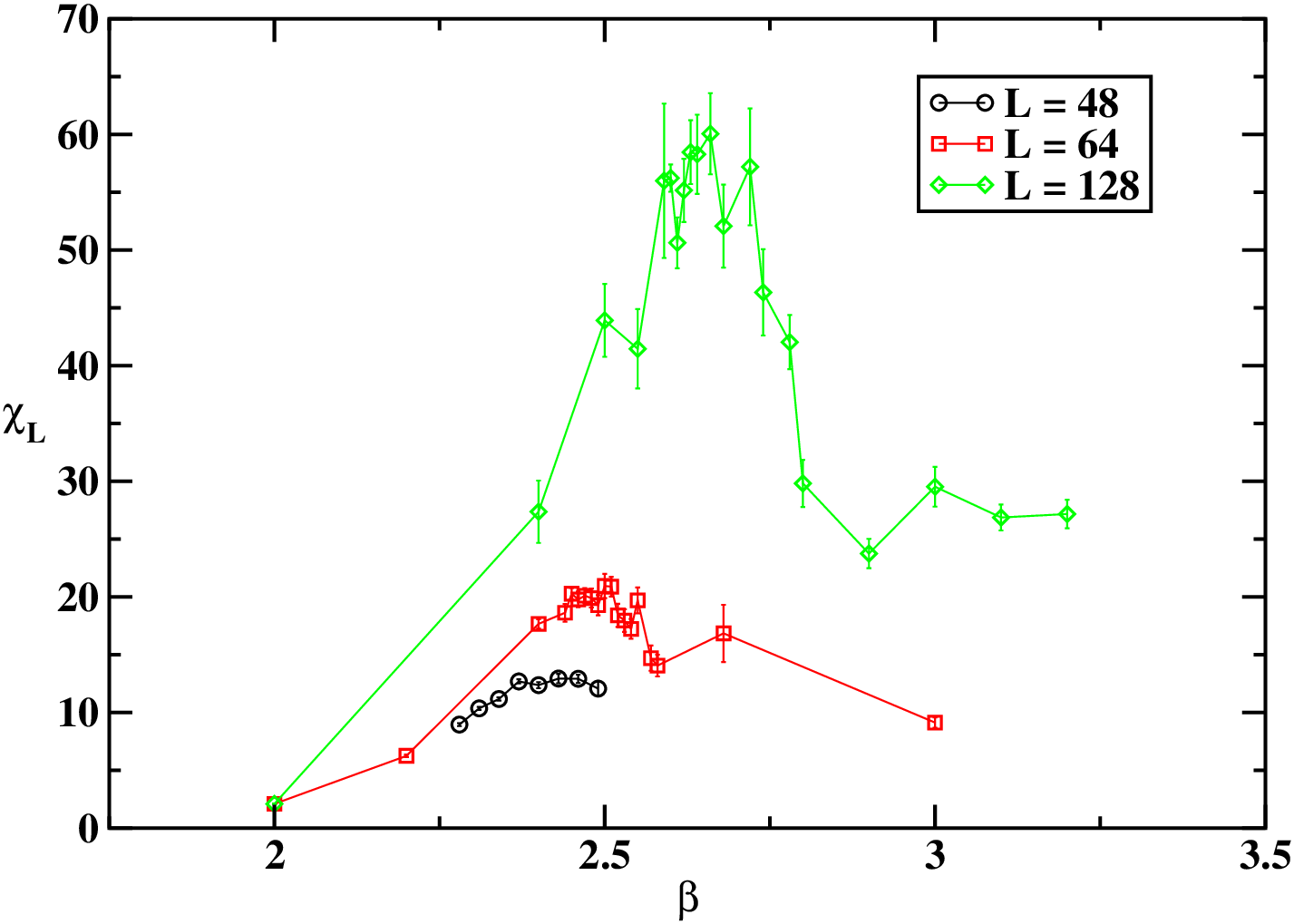}
\caption[]{Polyakov loop susceptibility $\chi_L$ on the lattices $L^2 \times 8$, 
with $L=48, 64, 128$.}
\label{susc_fig}
\end{figure}

At finite volume the transition manifests through a peak in the magnetic susceptibility of the 
Polyakov loop, defined as
\begin{equation}
\chi_L=L^2 (\langle |P|^2  \rangle -\langle |P| \rangle^2) \quad , \quad 
P = \frac{1}{L^2} \ \sum_x P(\vec{x}) \quad .
\end{equation}
The value of the coupling at which this happens is the pseudo-critical coupling, 
$\beta_{pc}$. By increasing the spatial volume, the position of the peak moves towards 
the (nonuniversal) infinite volume critical coupling, $\beta_c$. 
In Fig.~\ref{susc_fig} the behavior of $\chi_L$ around the transition is shown 
for $L=48, 64, 128$. The value of $\beta_{pc}$ for a given $L$ is determined by 
interpolating the values of the susceptibility $\chi_L$ around the peak by a 
Lorentzian function. In Table~\ref{beta_pc_tab} we summarize the resulting values of 
$\beta_{pc}$ and the peak values of the susceptibility $\chi_L$ for the several volumes 
considered in this work (we included also the determination for $L=32$, taken from the 
first paper in Ref.~\cite{chernodub}).

\begin{table}[ht]
\centering\caption[]{$\beta_{pc}$ and peak value of the Polyakov loop susceptibility 
$\chi_L$ on the lattices $L^2 \times 8$.}
\vspace{0.2cm}
\begin{tabular}{|c|l|c|}
\hline
 $L$ & $\beta_{pc}$ & $\chi_{L,\mbox{\scriptsize max}}$ \\
\hline
 32 & 2.346(2), Ref.~\cite{chernodub} & \\
 48 & 2.4238(67) & 12.93(41) \\
 64 & 2.4719(39) & 20.09(66) \\
 96 & 2.5648(96) & 38.8(1.6) \\
128 & 2.6526(59) & 60.1(3.5) \\
150 & 2.68(1)    & 92.6(8.0) \\
200 & 2.7336(69) & 144(12)   \\
256 & 2.7780(40) & 220(20)   \\
\hline
\end{tabular}
\label{beta_pc_tab}
\end{table}

In order to apply the finite size scaling (FSS) program, the location of the 
infinite volume critical coupling $\beta_c$ is needed. The scaling law by which 
$\beta_c$ can be extracted from the known values of $\beta_{pc}(L)$ depends on the 
nature of the transition and, in particular, on the behavior of the correlation length. 
There are, in principle, two hypotheses to be tested: first order and BKT transition.
The hypothesis of first order transition is not incompatible with data for the 
peak susceptibility for $L\geq 128$. However, the corresponding scaling law 
for the pseudo-critical couplings,
\begin{equation}
\beta_{pc}=\beta_c+\frac{A}{L^2} \quad ,
\label{b_pc_first}
\end{equation}
seems to be ruled out by our data ($\chi^2$/d.o.f equal to 5.6 for $L\geq 96$, 
3.7 for $L\geq 128$, 2.1 for $L\geq 96$).

Assuming the essential scaling of the BKT transition, {\it i.e.} $\xi \sim e^{bt^{-\nu}}$, 
the scaling law for $\beta_{pc}$ becomes
\begin{equation}
\beta_{pc}=\beta_c+\frac{A}{(\ln L + B)^{\frac{1}{\nu}}} \quad .
\label{b_pc}
\end{equation}
The index $\nu$ characterizes the universality class of the system. For example,
$\nu=1/2$ holds for the 2$d$ $XY$ universality class. 

We tried at first a 4-parameter fit of the data for $\beta_{pc}(L)$ given in 
Table~\ref{beta_pc_tab} with the law given in Eq.~(\ref{b_pc}). We excluded 
systematically from the fit the data for $\beta_{pc}(L)$ at the lowest spatial volumes,
looking for a region of stability of the parameters. Defining 
$L_{\mbox{\scriptsize min}}$ as the smallest value of $L$ for which $\beta_{pc}(L)$ 
has been considered in the fit, we could not find a stable fit for 
$L_{\mbox{\scriptsize min}} < 96$. In particular, we found that the $\chi^2$/d.o.f.
is $\approx 10$ for $L_{\mbox{\scriptsize min}} = 32$, $\approx 6$ for 
$L_{\mbox{\scriptsize min}} = 48$ and $\approx 1.8$ for $L_{\mbox{\scriptsize min}} = 64$.
Moreover, we observed a strong dependence of the fit parameters on the starting
values used in the MINUIT minimization code, although the resulting fitting curve turned 
out to be in general rather stable. The instability of parameters becomes less severe
when $L_{\mbox{\scriptsize min}}$ increases and, in particular, for 
$L_{\mbox{\scriptsize min}} = 64$ the parameter $\nu$ becomes compatible
with the $XY$ value, $\nu=0.5$, although still undergoing large fluctuations under 
change of the starting conditions of the MINUIT minimization procedure. A stable
fit could be achieved only for $L_{\mbox{\scriptsize min}}=96$ and these are the
resulting parameters:
\[
\beta_c=3.06(16)\;, \;\;\;\;\;  A=-5.3(5.1)\; \;\;\;\;\;
B=-1.4(1.0) \;, \;\;\;\;\; \nu=0.49(16) \;\;\;\;\; (\chi^2/\mbox{d.o.f.}=1.5).
\]

We repeated then the fit with the law~(\ref{b_pc}) keeping the parameter $\nu$ fixed
at the $XY$ value, $\nu= 1/2$, thus reducing to three the number of free parameters in the
fit. In this case, the fit instability is highly suppressed with respect to the 
previous 4-parameter analysis and, indeed, already for $L_{\mbox {\scriptsize min}}\geq 
48$ we can quote stable values of the fit parameters (see Table~\ref{3_param_fit}). 
One can see that an acceptable $\chi^2$/d.o.f. and a stable fit are obtained for 
$L_{\mbox {\scriptsize min}}=64$ and $L_{\mbox {\scriptsize min}}=96$ and that,
for the latter volume, $\beta_c$ is consistent with the result of the 4-parameter fit. 
We take therefore $\beta_c=3.06(11)$ as our estimation for the infinite volume
critical coupling. The determination of $\beta_c$ is the first main result of this work.

\begin{figure}[tb]
\centering
\includegraphics[width=15cm]{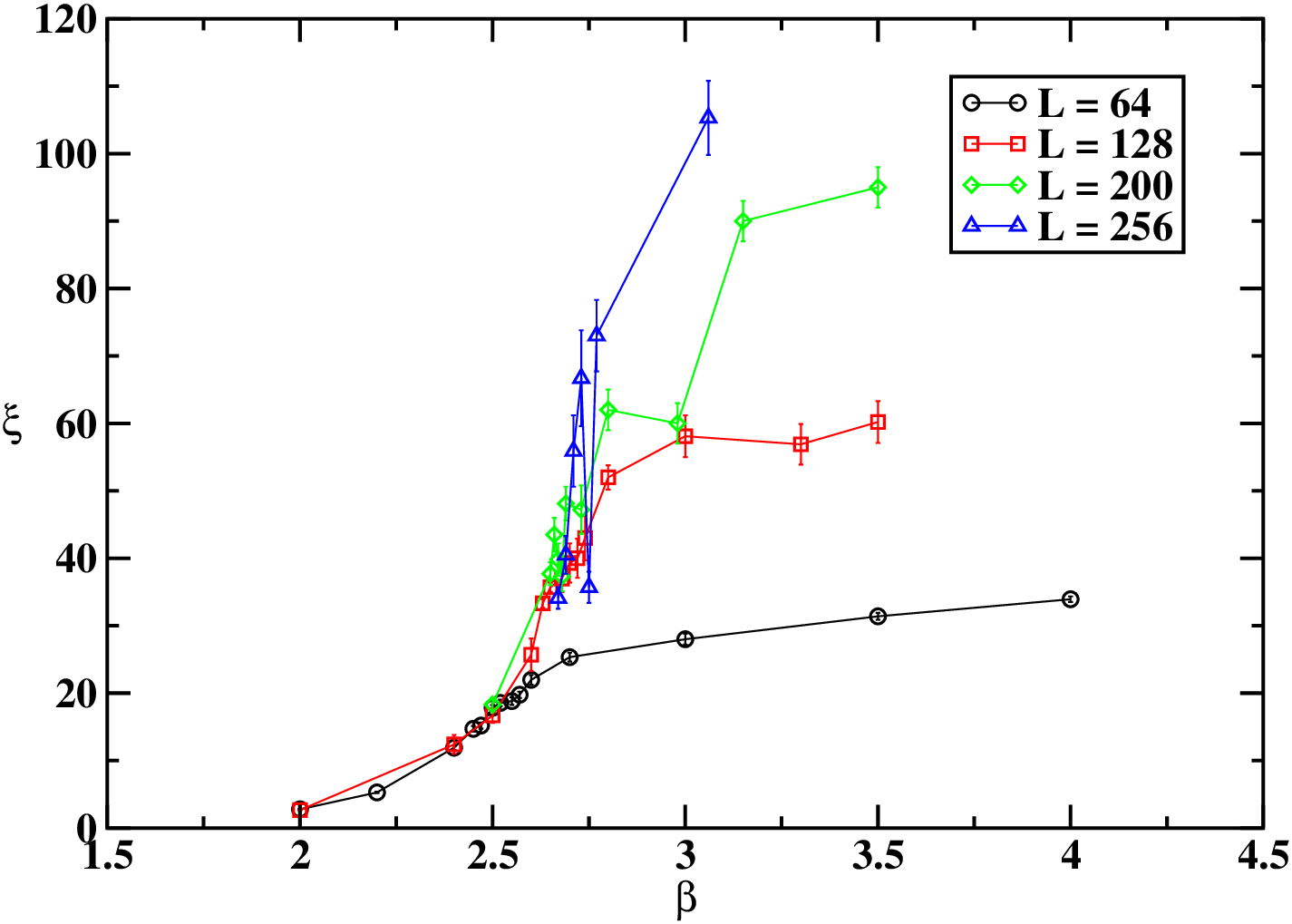}
\caption[]{Correlation length {\it vs} $\beta$ on the lattices $L^2 \times 8$, 
with $L$=64, 128, 200, 256. The correlation length is determined {\it as if} the 
exponential fall-off with distance of the 2-point correlator of the Polyakov loop apply
for all $\beta$ values.}
\label{xi_fig}
\end{figure}

While performing this work, we considered also the possibility to extract the index $\nu$
from directly fitting the essential scaling law $\xi \sim e^{bt^{-\nu}}$ against
lattice data for the correlation length taken for several $\beta$ values and for several 
volumes. To be more precise, for each considered value of $L$ and for several $\beta$
values across $\beta_{pc}(L)$, we determined the correlation length $\xi_L$ 
as the inverse decay length of the 2-point correlator of the Polyakov loops, 
interpolating the latter {\it as if} the exponential fall-off with the distance 
apply even above $\beta_{pc}$, where, in fact, this correlator decays power-like 
(see Fig.~\ref{xi_fig}). 
At each volume it happens that $\xi_L$ defined in this way increases with $\beta$ till 
$\beta \approx \beta_{pc}$ and then saturates, consistently with the fact that the region
of power-like behavior has been reached. It occurs, however, that the set of all lattice 
data for the correlation length $\xi_L$ that, at each volume, belong to the region 
$\beta < \beta_{pc}(L)$, lie approximately on the same curve.
This is expected to occur more and more accurately as the thermodynamic limit 
is approached. One could then try to fit the lattice data for $\xi_L$ falling
on this curve with the essential scaling law and extract $\nu$. Unfortunately, the 
quality of our data did not allow us to have a stable fit and we had to reject this 
method. It cannot be excluded, however, that it will be reconsidered in possible future 
studies of the same kind.

\begin{table}[ht]
\centering\caption[]{Results of the 3-parameter fit of the values of $\beta_{pc}(L)$ 
with the law~(\ref{b_pc}), $\nu=1/2$ fixed.}
\vspace{0.2cm}
\begin{tabular}{|c|l|c|c|c|}
\hline
 $L_{\mbox{\scriptsize min}}$ & $\beta_{c}$ & $A$ & $B$ &  $\chi^2$/d.o.f.  \\
\hline
32 & 5.103(50) &  -1523(86) &  20.03(48) &  13.  \\
48 & 4.65(23)  &  -699(239) &  13.8(2.1) &  4.3  \\
64 & 3.44(15)  &  -42(24)   &  2.4(1.4)  &  1.2  \\
96 & 3.06(11)  &  -4.7(4.3) &  -1.5(1.1) &  0.76 \\
\hline
\end{tabular}
\label{3_param_fit}
\end{table}

Once an estimation for $\beta_c$ has been achieved, we can use the FSS 
analysis, which holds just at $\beta_c$, to extract other critical indices. 
An interesting example is the magnetic critical index, $\eta$, which enters
the scaling law 
\begin{equation}
\chi_L(\beta_c) \sim L^{2-\eta} \quad .
\label{eta_scale}
\end{equation}
Actually in this law one should consider logarithmic corrections 
(see~\cite{Kenna-Irving,Hasenbusch} and references therein) and, indeed,
recent works on the $XY$ universality class generally include them. However,
taking these corrections into account for extracting critical indices calls
for very large lattices even in the $XY$ model; for the theory under 
consideration to be computationally tractable, we have no choice but to neglect
logarithmic corrections.

Setting the coupling $\beta$ at the value of our best estimation for $\beta_c$,
i.e. $\beta=3.06$, we determined the susceptibilities $\chi_L(\beta_c)$ for 
several volumes (see Table~\ref{susc_b_c} for the results). Then, following FSS,
we fitted the results with the law $\chi_L(\beta_c)=A L^{2-\eta}$ and got
\begin{equation} 
A=0.0171(10) , \quad \eta=0.496(15) \quad (\chi^2/\mbox{d.o.f.}=0.60) \quad .
\end{equation} 

This is the second main result of our paper. We stress that this value for
$\eta$ is by far incompatible with the 2$d$ $XY$ value, $\eta_{XY}=0.25$. 
The most extreme consequence of this finding is that the deconfinement transition in 
the 3$d$ $U(1)$ LGT at finite temperature does not belong to the same universality class 
as 2$d$ $XY$ spin model. This would contradict the Svetitsky-Yaffe conjecture, raising 
a problem in the understanding of the deconfinement mechanism in gauge theories. We will
further comment on this issue in the next section, discussing possible ways out.

\begin{table}[ht]
\centering
\caption[]{Values of $\chi_L(\beta_c=3.06)$ on the lattices $L^2 \times 8$.}
\vspace{0.2cm}
\begin{tabular}{|c|c|}
\hline
$L$ & $\chi_L(\beta_c=3.06)$ \\
\hline
48  &  5.732(42) \\
64  &  8.887(76) \\
96  & 17.16(95)  \\
128 & 25.37(60)  \\
150 & 31.52(75)  \\
200 & 50.1(2.5)  \\
256 & 65.9(4.6)  \\
\hline
\end{tabular}
\label{susc_b_c}
\end{table}

In such a situation, it becomes particularly useful to have another determination of the 
index $\eta$, by an independent approach. Following the strategy of our previous paper~\cite{beta_szero}, we define an {\em effective} $\eta$ index, through
the 2-point correlator of Polyakov loops, according to
\begin{equation}
\eta_{\mbox{\scriptsize eff}}(R) \equiv \frac{\log [\Gamma (R)/\Gamma (R_0)]}
{\log [R_0/R]} \quad ,
\label{eff_eta_def}
\end{equation}
with $R_0$ chosen equal to 10, as in Ref.~\cite{beta_szero}. This quantity is 
constructed in such a way that it exhibits a {\it plateau} in $R$ if the correlator 
obeys the law~(\ref{PLhight}), valid in the deconfined phase. 

\begin{figure}[tb]
\centering
\includegraphics[width=15cm]{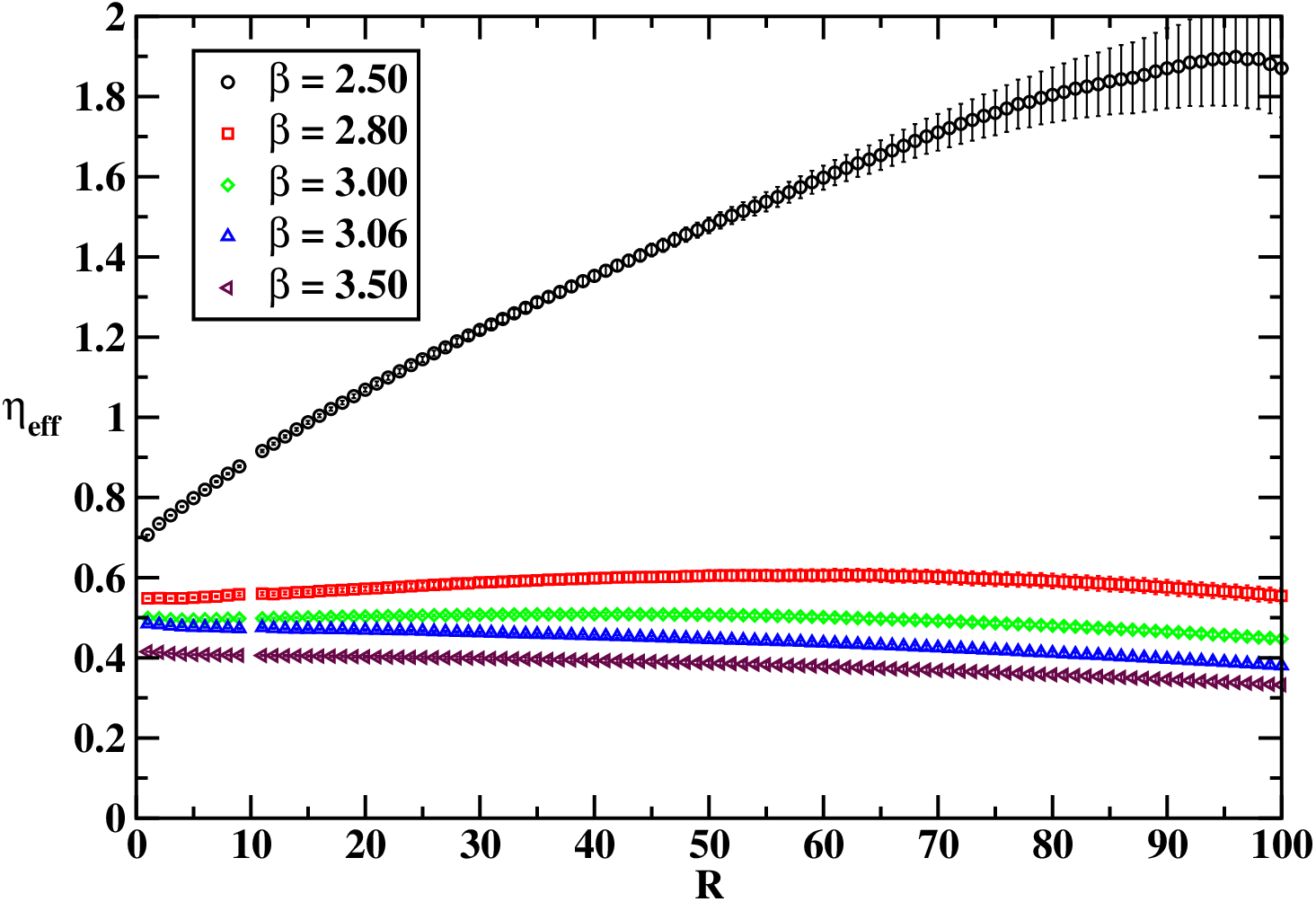}
\caption[]{$\eta_{\mbox{\scriptsize eff}}$ {\it vs} $R$ on the lattice $200^2\times 8$ 
for several $\beta$ values.}
\label{eta_eff_200}
\end{figure}

In Fig.~\ref{eta_eff_200} we present $\eta_{\mbox{\scriptsize eff}}$ as a function of 
the distance $R$ for several $\beta$ values on the lattice with $L=200$. A drastic change 
in the behavior of $\eta_{\mbox{\scriptsize eff}}$ is observed across the value 
$\beta^* \approx 3$. In particular for $\beta > \beta^*$ a plateau develops at 
short distances, deviations at large $R$ being interpreted as a finite volume effect 
which becomes stronger with increasing $\beta$ since $\xi$ diverges in the deconfined 
phase. The appearance of this plateau is an indication that the correlator takes the 
power behavior expected for a BKT transition.

\begin{figure}[tb]
\centering
\includegraphics[width=15cm]{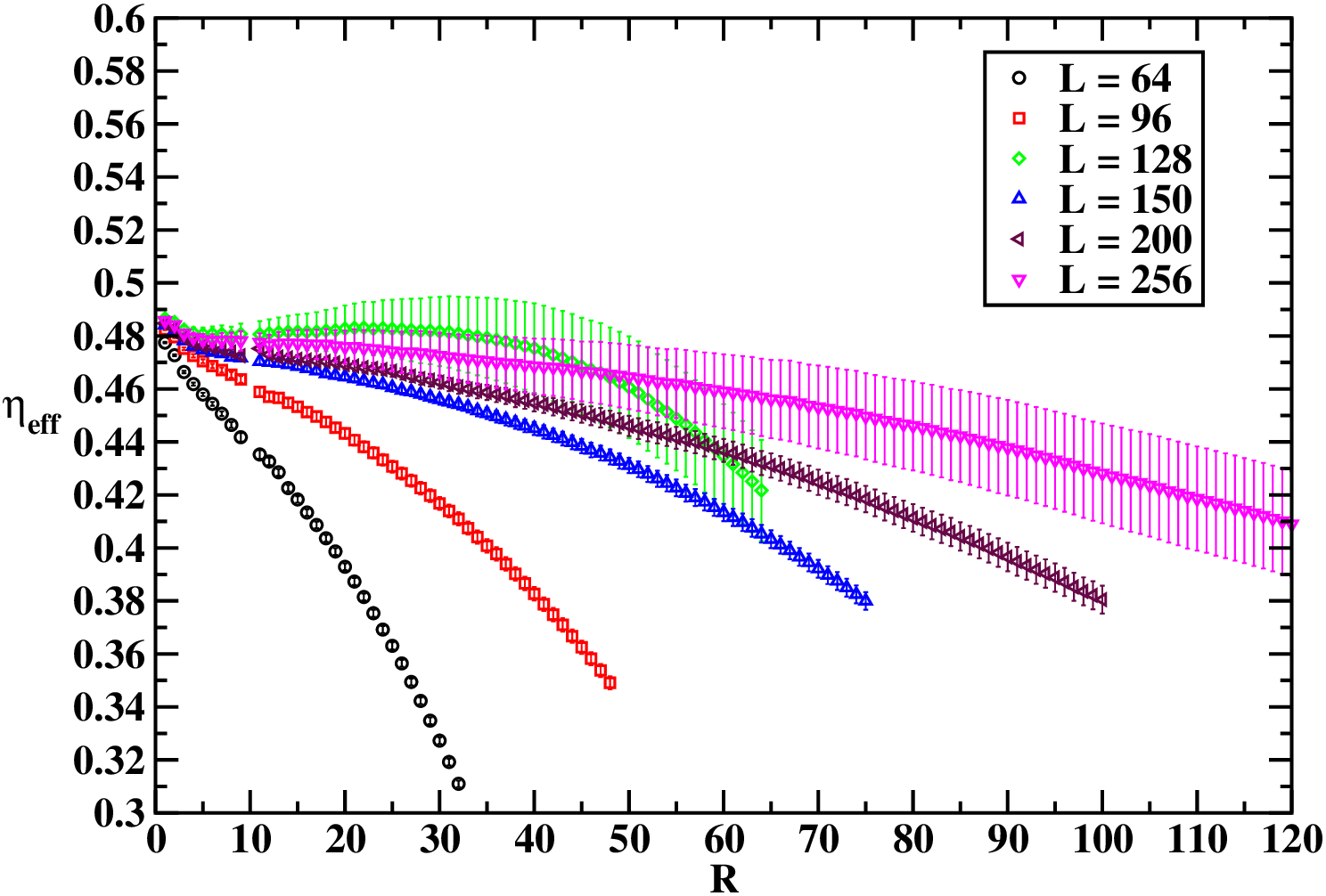}
\caption[]{$\eta_{\mbox{\scriptsize eff}}$ {\it vs} $R$ at $\beta_c$ on lattices with 
several values of $L$.}
\label{eta_eff_3.06}
\end{figure}

The analysis of the behavior of $\eta_{\mbox{\scriptsize eff}}(R)$ has been
repeated setting $\beta$ at our estimated value for $\beta_c$, {\it i.e.} $\beta=3.06$, 
and increasing the spatial extent of the lattice. It turns out (see 
Fig.~\ref{eta_eff_3.06}) that a plateau develops at small distances when $L$ increases
and that the extension of this plateau gets larger with $L$, consistently with 
the fact that finite volume effects are becoming less important. The plateau value of 
$\eta_{\mbox{\scriptsize eff}}$ can be estimated as 
$\eta_{\mbox{\scriptsize eff}}(R=6)$ on the 256$^2\times 8$ lattice and is equal 
to 0.4782(25); it agrees with our previous determination of the index $\eta$.

\begin{figure}[tb]
\centering
\includegraphics[width=15cm]{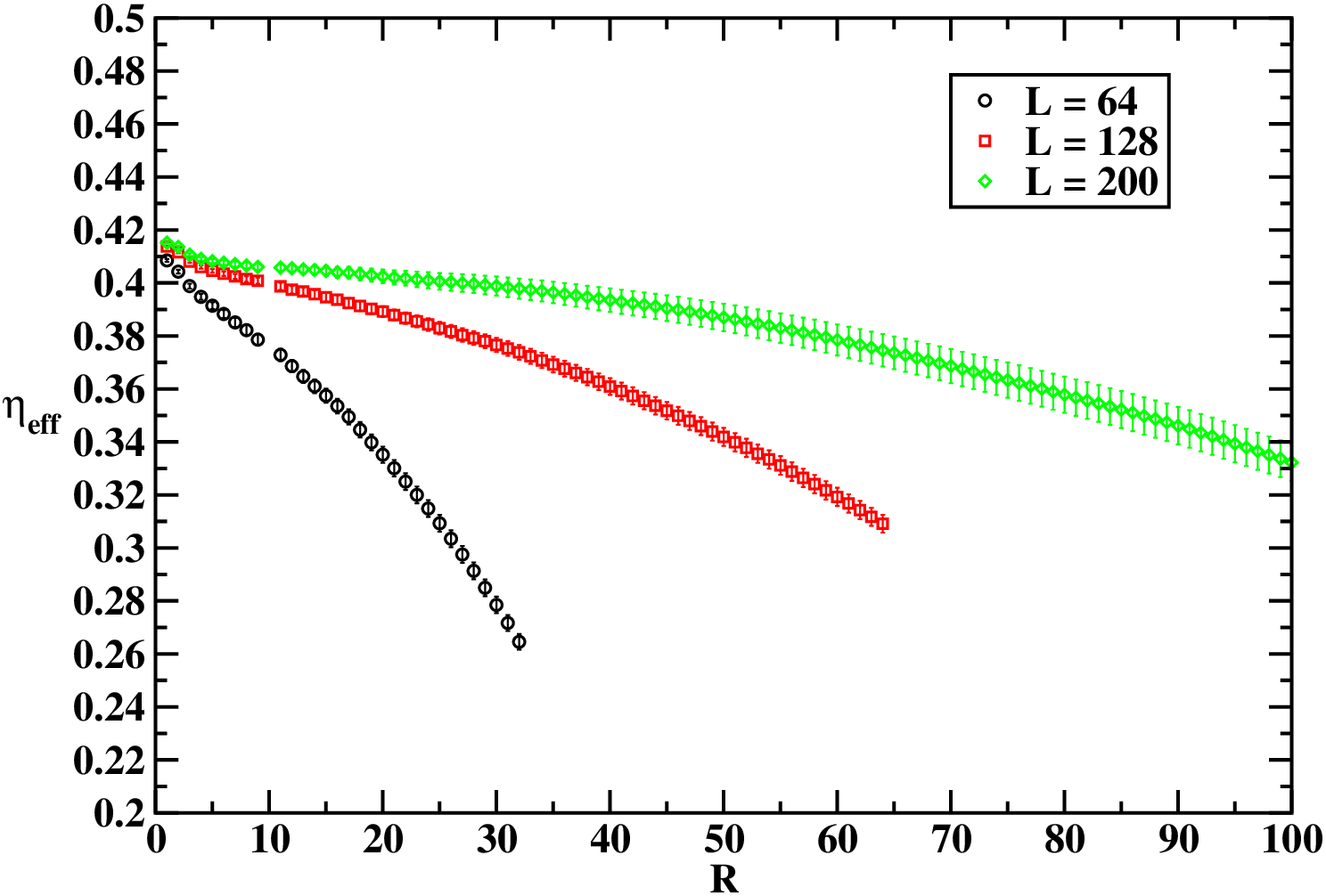}
\caption[]{$\eta_{\mbox{\scriptsize eff}}$ {\it vs} $R$ at $\beta=3.50$ on lattices with 
several values of $L$.}
\label{eta_eff_3.50}
\end{figure}

The scenario described by Fig.~\ref{eta_eff_3.06} for $\beta=\beta_c$ must 
be valid for any $\beta > \beta_c$, if the system undergoes a BKT transition, since
the correlator must obey a power law in the whole high-$\beta$ phase. 
We have found that this is indeed the case by performing an analysis similar to that 
shown in Fig.~\ref{eta_eff_3.06} at several $\beta$ values larger than $\beta_c$
(see Fig.~\ref{eta_eff_3.50} for the case of $\beta=3.50$, which leads to 
$\eta\approx 0.41$)).
We observe that, in general, $\eta(\beta) < \eta(\beta_c)$ for $\beta>\beta_c$ and had 
we estimated $\beta_c$ from the 3-parameter fit with $L_{\mbox{\scriptsize min}}=64$, 
instead of that with $L_{\mbox{\scriptsize min}}=96$, {\it i.e.} 3.44(15) instead of 
3.06(11), the resulting $\eta$ would change by little and keep still much larger
than the $XY$ value 0.25.

\section{Discussions}

In this paper we have studied the critical behavior of the $3d$ $U(1)$ LGT 
at finite temperature. We worked on isotropic lattice with the temporal extension 
$N_t=8$. The pseudo-critical coupling was determined through the peak in 
the susceptibility of the Polyakov loop. The infinite-volume critical coupling 
has then been computed assuming the scaling behavior of the form (\ref{b_pc}).
Our fitting gives the value $\beta_c=3.06(11)$. The deconfinement phase is the phase 
where $\beta\geq\beta_c$. The detailed study of the deconfined phase is clearly 
beyond the scope of the present paper. Nevertheless, this finding has immediate 
impact on the previous studies of the model. A thorough investigation of 
the deconfinement phase was performed in Ref.~\cite{chernodub}. However, all 
$\beta$-values used there are smaller than the infinite-volume critical coupling.
When the thermodynamic limit is approached the critical coupling increases so
that the numerical results of Ref.~\cite{chernodub} would refer rather to the
confinement phase of the infinite-volume theory.
This is indeed the case as we explained in the previous section. One sees 
from Fig.~\ref{xi_fig} that the correlation length (inverse of the string tension) 
grows till the pseudo-critical value of $\beta$ is reached. Therefore, the string 
tension is non-vanishing for all values of $\beta$ used in Ref.~\cite{chernodub}. 
We conclude the much larger $\beta$-values are needed than those used in 
Ref.~\cite{chernodub} to really probe the physics of the deconfinement phase in 
the large volume limit. 
This however might call for very large lattice sizes so the feasibility of such 
a study is not clear at present. 

Furthermore, the index $\nu$ has been extracted from the scaling of the 
pseudo-critical couplings (\ref{b_pc}). Its value does agree well with the expected 
$XY$ value $\nu=1/2$. Of course, it would be desirable to extract this index directly from 
the correlation of the Polyakov loops along the line described in the previous section.

One of the main results of our paper is the computation of the index $\eta$ 
which turns out to be $\eta \approx 0.496$. This value is essentially larger than 
expected and requires some discussion. The easiest explanation would be to state that 
the spatial lattice size used ($L\in [32-256]$) is still too small to exhibit 
the correct scaling behavior, hence the wrong values for $\beta_c$ and $\eta$ 
follow. However, if one makes a plot of $\beta_{pc}(L)$ vs $L$, one can see, by looking 
at the trend of data, that it is unlikely that $\beta_c$ is much larger than our estimate.
In fact, our fits with the scaling law (\ref{b_pc}) show that $\beta_c$ decreases
when $L_{\mbox{\scriptsize min}}$ increases (see Table~\ref{3_param_fit}). Therefore,
our result is most likely an overestimation. This implies that the true $\eta$ is likely even
larger than what we found. In any case, even if we use for $\beta_c$  rather unlikely
value $\beta_c=3.50$, Fig.~\ref{eta_eff_3.50} suggests that $\eta\approx 0.41$, much
above the $XY$ value. 

The next objection against our result could be the fact that we have neglected 
logarithmic corrections to the scaling law (\ref{eta_scale}). It looks for us rather 
strange that logarithmic corrections can lead to decreasing $\eta$ almost by two times.
We want to mention that including naively such corrections into our fits always results 
in the increasing of $\eta$ values, though these values are unstable against the 
maximal lattice size included into fit. 
Thus, although we cannot rule out this possibility, we do not think 
that neglecting logarithmic corrections results in such a wrong prediction for $\eta$.

Let us give a simple argument why the index $\eta$ can be different from its $XY$ value.
Consider the anisotropic lattice. We would like to study the limit of large $\beta_s$. 
In the limit $\beta_s=\infty$ the spatial plaquettes are frozen 
to unity. That means, the ground state is a state where all spatial fields are 
pure gauge, i.e. $U_n(x) \ = \ V_x V^*_{x+e_n}  \ , \ n=1,2$. Perform now a change 
of variables $U_0(x)\to V_x U_0(x)V^*_{x+e_0}$. Then it is easy to see 
that in the leading order of the large-$\beta_s$ expansion the partition function 
factorizes into the product of $N_t$ independent $2d$ $XY$ models.
Let us now look at the correlations of the Polyakov loops. 
Since the Polyakov loop is the product of gauge fields in the temporal direction, 
the correlation function factorizes, too, and becomes a product of independent 
$XY$ correlations, i.e. 
\begin{equation}
\Gamma_{U(1)}(\beta_s=\infty , \beta_t) \ = \ \left [ \Gamma_{XY}(\beta_t) \right ]^{N_t} \ .
\label{PLcorr}
\end{equation}
Hence, for asymptotically large $R\gg 1$, we get
\begin{equation}
\Gamma_{U(1)}(\beta_s=\infty , \beta_t\geq\beta_t^{cr}) \ 
\asymp \ \left [ \frac{1}{R^{\eta_{XY}}} \right ]^{N_t} \ .
\label{PLcorr1}
\end{equation}
This leads to a simple relation
\begin{equation}
\eta (\beta_s=\infty,\beta_t^{cr}) \ = \ N_t \ \eta_{XY} \ .
\label{etabeta_sinfty}
\end{equation}

\begin{figure}[tb]
\centering
\includegraphics[width=15cm]{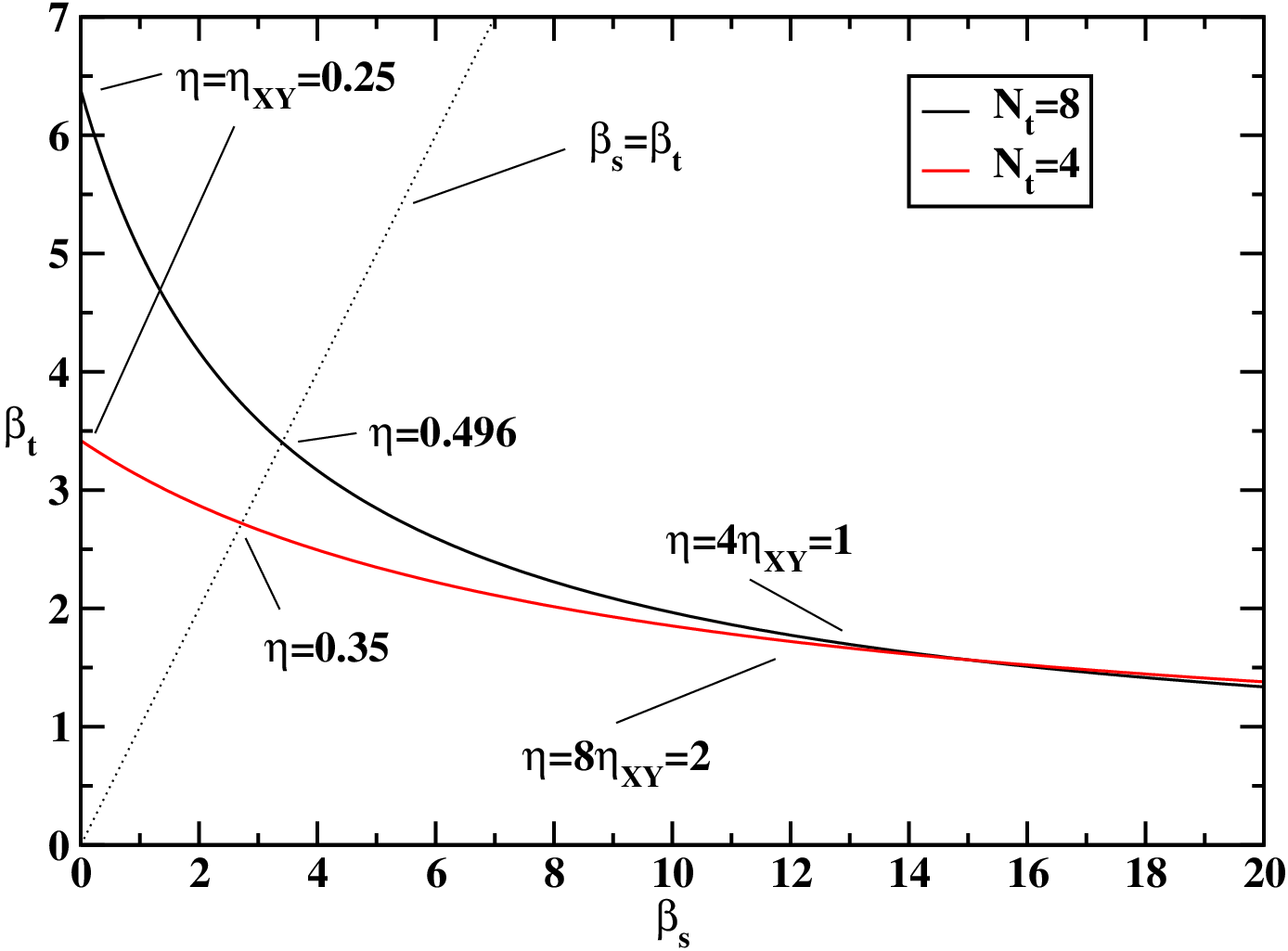}
\caption[]{Conjectured behavior of $\eta$ in the $\beta_s$-$\beta_t$ plane. 
Results for $N_t=4$ are taken from Ref.~\cite{bazavov}.}
\label{eta_phase}
\end{figure}

Some conclusions could now be drawn. The critical behavior of the $3d$ $U(1)$ 
LGT in the limit $\beta_s\to\infty$ is also governed by the $2d$ $XY$ model. 
Nevertheless, the effective index $\eta$ appears to be $N_t$ times of its $XY$ value. 
Now, for $\beta_s=0$ we have $\eta (\beta_s=0,\beta_t^{cr}) \ = \  \eta_{XY}$. 
This relation and formula (\ref{etabeta_sinfty}) allow to conjecture that
\begin{equation}
\eta_{XY} \ \leq \ \eta (\beta_s,\beta_t^{cr}) \ \leq \ N_t \ \eta_{XY} \ .
\label{etabeta_s}
\end{equation}
$\beta_s=0$ corresponds to the lower limit while $\beta_s=\infty$ corresponds to the
upper limit. 
In general, $\eta$ could interpolate between two limits with $\beta_s$. 
Whether this interpolation is monotonic or there exists critical value $\beta_s^{cr}$,
such that $\eta (\beta_s\leq \beta_s^{cr},\beta_t^{cr}) \ = \ \eta_{XY}$ and $\eta$ 
changes monotonically above $\beta_s^{cr}$, cannot be answered with data we have and 
requires simulations on the anisotropic lattices. In the paper~\cite{rgu1} 
a renormalization group study of $3d$ $U(1)$ model at small $\beta_s$ will
be presented and computations of the leading correction to the large 
$\beta_s$ behavior will be given. The results of our computations 
support the scenario that the index $\eta$ depends on the ratio $\beta_s/\beta_t$. 
Recently, we have obtained the results of simulations for $N_t=2,4$ performed 
by A.~Bazavov~\cite{bazavov}. His results also point in the direction of our scenario 
(see Fig.~\ref{eta_phase}). In Fig.~\ref{eta_phase} we plot a possible behavior 
of $\eta$ supposing the monotonic dependence. 

Finally, it is worth mentioning that the factorization in the large $\beta_s$ 
limit does not affect the index $\nu$. It follows from its definition 
(\ref{PLlowt}) that in this limit $\nu=1/2$ as in the $XY$ model. 
We expect therefore that $\nu$ equals $1/2$ for all $\beta_s$ and is thus universal.

In view of our results it might be worth to perform numerical simulations 
for small but nonvanishing $\beta_s$ and for larger volumes.  
The feasibility of a study with larger volumes and better accuracy relies strongly 
on the possibility to improve the simulation code. Promising directions could be 
simulations of the dual formulation of the model (possibly with a cluster algorithm) 
or the use of the L\"uscher-Weisz algorithm~\cite{LW}. 
The development of these directions is in progress.

\vspace{1.0cm} \noindent
{\Large \bf Acknowledgment} \vspace{0.5cm}

O.B. thanks for warm hospitality the Dipartimento di Fisica dell'Universit\`a 
della Calabria and the INFN Gruppo Collegato di Cosenza during the work on 
this paper. Numerical simulations were performed on the linux PC farm ``Majorana'' of 
the INFN-Cosenza and on the GRID cluster at the ITP-Kiev. Authors would like to thank 
A.~Bazavov for interesting discussions and for providing us with the results of his 
simulations on the lattices with $N_t=2,4$ prior to publication.

\end{document}